\documentclass[a4paper,fleqn]{cas-dc}

\usepackage[authoryear]{natbib}

\usepackage{hyperref}
\usepackage{amssymb}
\usepackage{latexsym}
\usepackage{url}
\usepackage{xcolor}
\usepackage{mathtools}
\usepackage{adjustbox}
\usepackage{float}

\DeclareMathOperator*{\OPLUS}{\oplus}

\def\tsc#1{\csdef{#1}{\textsc{\lowercase{#1}}\xspace}}
\tsc{WGM}
\tsc{QE}
\begin{document}
\let\WriteBookmarks\relax
\def\floatpagepagefraction{1}
\def\textpagefraction{.001}

% % Short title
\shorttitle{D-STGCNT for Assessment of Patient Physical Rehabilitation}    

% % Short author
% \shortauthors{<short author list for running head>}  

\title [mode = title]{D-STGCNT: A Dense Spatio-Temporal Graph Conv-GRU Network based on Transformer for Assessment of Patient Physical Rehabilitation}  

% \tnotemark[<tnote number>] 



\author[1]{Youssef MOURCHID}[]

\cormark[1]

\fnmark[1]

\ead{ymourchid@cesi.fr}


% \credit{<Credit authorship details>}

\affiliation[1]{organization={CESI LINEACT},
            addressline={UR 7527}, 
            city={Dijon},
%          citysep={}, % Uncomment if no comma needed between city and postcode
            postcode={21800}, 
            state={},
            country={France}}

\author[2]{ Rim SLAMA}[]

% % Footnote of the second author
\fnmark[1]

% Email id of the second author

% % URL of the second author

% % Credit authorship

\affiliation[2]{organization={CESI LINEACT},
            addressline={UR 7527}, 
            city={Lyon},
%          citysep={}, % Uncomment if no comma needed between city and postcode
            postcode={69100}, 
            state={},
            country={France}}

\ead{rsalmi@cesi.fr}

\cortext[1]{Corresponding author}

\fntext[1]{These authors contributed equally to this work.}


\begin{abstract}
 This paper tackles the challenge of automatically assessing physical rehabilitation exercises for patients who perform the exercises without clinician supervision. The objective is to provide a quality score to ensure correct performance and achieve desired results. To achieve this goal, a new graph-based model, the Dense Spatio-Temporal Graph Conv-GRU Network with Transformer, is introduced. This model combines a modified version of STGCN and transformer architectures for efficient handling of spatio-temporal data. The key idea is to consider skeleton data respecting its non-linear structure as a graph and detecting joints playing the main role in each rehabilitation exercise. Dense connections and GRU mechanisms are used to rapidly process large 3D skeleton inputs and effectively model temporal dynamics. The transformer encoder's attention mechanism focuses on relevant parts of the input sequence, making it useful for evaluating rehabilitation exercises. The evaluation of our proposed approach on the KIMORE and UI-PRMD datasets highlighted its potential, surpassing state-of-the-art methods in terms of accuracy and computational time. This resulted in faster and more accurate learning and assessment of rehabilitation exercises. Additionally, our model provides valuable feedback through qualitative illustrations, effectively highlighting the significance of joints in specific exercises.

\end{abstract}
% % Use if graphical abstract is present
% \begin{figure}[ht]
% \centering
% \includegraphics[width=15cm,height=11.5cm]{Flowchart4.pdf}
% \caption{Flowchart of the proposed approach.}
% \label{fig:flowchart}
% \end{figure}

% % Research highlights
% \begin{highlights}
%  \item We propose a new method for evaluating physical rehabilitation exercises performed by patients without clinician supervision, using a Dense Spatio-Temporal Graph Conv-GRU Network with Transformer. 
 
%  \item The method uses skeleton data as a graph, detects key joints, and uses dense connections and GRU mechanisms for processing 3D skeleton inputs
 
%  \item The transformer encoder's attention mechanism is employed to focus on relevant parts of the input skeleton sequence. 
 
% \item The method is tested on KIMORE and UI-PRMD datasets and showed its potential by outperforming the current state-of-the-art methods and giving a reliable feedback.
% \end{highlights}

% % Keywords
% % Each keyword is seperated by \sep

\maketitle

\section{Introduction}\label{sec1}

In the healthcare field, physical rehabilitation exercises play a crucial role in post-surgery recovery and managing various musculoskeletal issues \citep{thiry2022machine}. These exercises are usually monitored by a clinician in a hospital or clinic setting, but patients receive limited supervised sessions due to high expenses or staff availability problems. To achieve optimal recovery, it is vital that patients continue to perform the prescribed exercises correctly in their own homes. Recently advanced motion sensors were developed for capturing human motion \citep{alarcon2020upper}. In particular, low-cost vision depth cameras that are recently commercialized such as the Kinect vision device \citep{scott2022healthcare}. This latter are marker-less motion capture system using the time-of-flight (ToF) principle and is able to capture precisely RGB, depth images, and joint skeletal coordinates. 

In the context of patient rehabilitation, many works consider these joints for human motion analysis \citep{devanne2017multi,deb2022graph}. They are encouraged by their effectiveness shown in various action recognition applications \citep{yue2022action}. Motivated by this, our work aims to build an automatic model for physical rehabilitation exercise assessment using joint skeletal data of exercises as an input. The proposed model will help patients to continue their exercises independently while getting a feedback helping them improve the accuracy of their movements.

\begin{figure*}[ht]
\centering
\includegraphics[width=14cm,height=6cm]{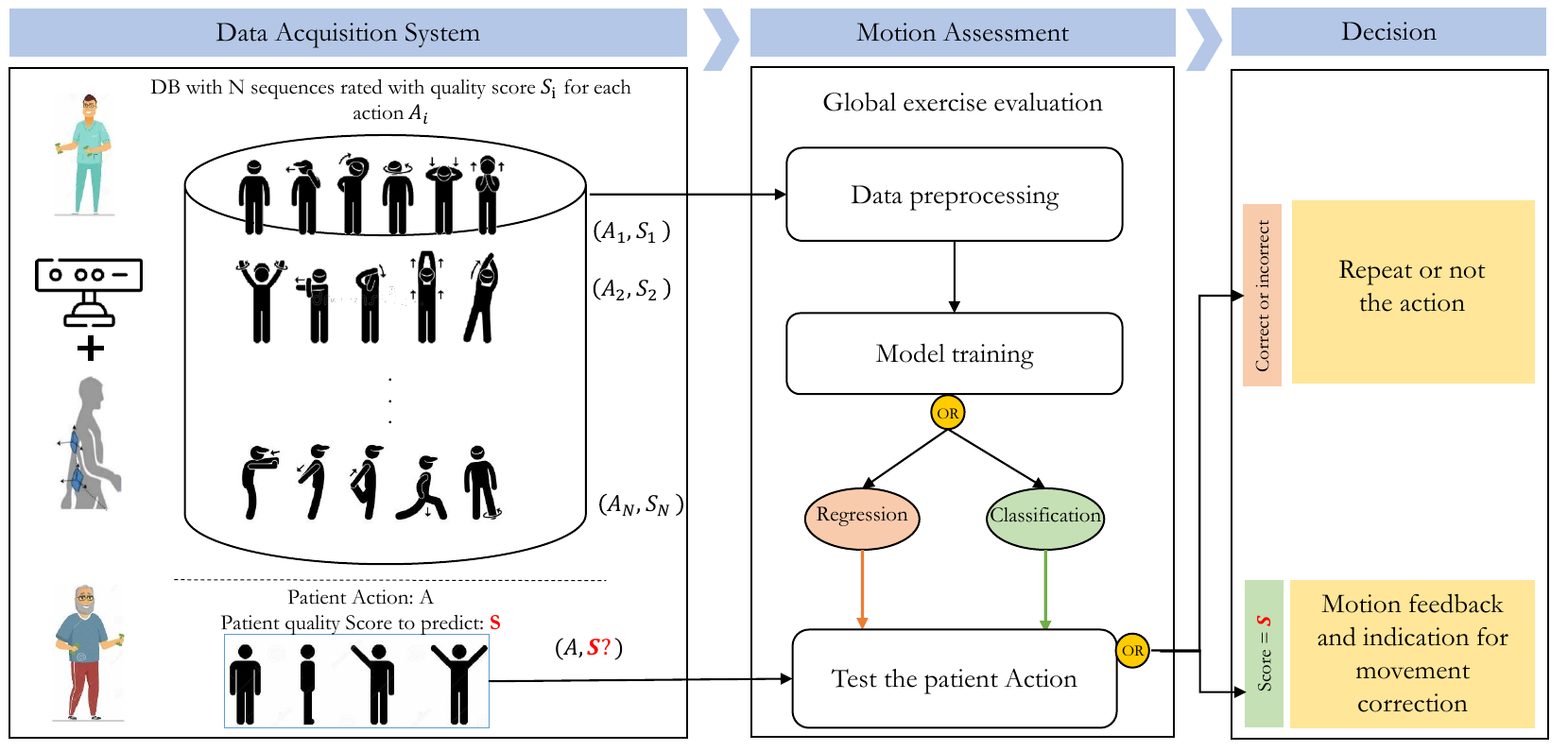}
\caption{Physical rehabilitation exercises process overview. } 
\label{fig:RehabilitationOverview}
\end{figure*}

In the literature, previous studies of \cite{hamaguchi2020support,pogorelc2012automatic} considered exercise evaluation as a binary classification (correct or incorrect) without being able to give a feedback for each exercise performance (see Figure ~\ref{fig:RehabilitationOverview}). Other approaches \cite{lee2019learning} predict a continuous score by addressing a regression problem and relying on handcrafted features (projected trajectory, relative trajectory, etc.), which mostly require time-consuming pre-processing and expert knowledge. Advancements in computer vision, driven by graphs, statistical techniques, and deep learning, have greatly enhanced visual data processing, which is particularly beneficial for improving rehabilitation exercises assessment \cite{mourchid2016image,benallal2022new,mourchid2021automatic,mourchid2023mr}.

Recently,  \cite{liao2020deep} leveraged the power of deep learning techniques for feature extraction by using deep spatio-temporal neural network model for outputting movement quality scores. Before feeding the network by input videos, they convert the latter to a fixed length. Nevertheless, these methods do not respect the topological structure of the skeleton and do not consider interaction among neighborhood joints. Recently, graphs have been extensively employed for various computer vision applications \citep{lafhel2021movie, mourchid2019movienet} and more particularly for skeleton-based action identification  since the human skeleton and a graph are comparable. Spatio-Temporal Graph Convolutional Networks (STGCN), a subcategory of Graph Convolutional Networks (GCN), was applied to skeleton-based activity recognition in \citep{yan2018spatial} by creating a spatio-temporal graph through the connection of detected joints of a human body in consecutive time steps.
Besides, one of the most significant deep learning developments over the past few years has been the Transformer architecture \citep{vaswani2017attention}. Beyond NLP, a variety of tasks, including image classification, image super-resolution, speech recognition, and particularly human motion analysis \citep{plizzari2021skeleton,zhang2023vit}, have shown that multi-head self-attention is effective. It is increasingly used to improve model accuracy by combining attention mechanisms with other deep learning blocks.

Inspired by the recent development and achievements of (GCNs) \citep{ahmad2021graph} and the observed effectiveness of transformers, we propose, in this work, an extended architecture based on GCNs coupled with the power of attention mechanism. The objective is to evaluate patient actions using sequential skeleton data. First, we use dense connections between STGC-GRU blocks which allow a more direct and efficient flow of information. Dense connections have been shown to improve results for our assessment task making it easier for the network to learn complex features and patterns \citep{huang2017densely}. They allow to alleviate the vanishing-gradient problem, strengthen feature propagation and encourage feature reuse at different scales, which leads to better performance.

Second, instead of using Convolutional Long Short Term Memory (ConvLSTM) layers as in \cite{deb2022graph}, we propose to employ Convolutional Gated recurrent units (ConvGRU). The main advantage of ConvGRU over ConvLSTM is their simpler structure, which makes them more computationally efficient. ConvGRU only has two gates (an update gate and a reset gate) compared to ConvLSTM, which has three (an input gate, an output gate, and a forget gate). This simpler structure allows ConvGRU to have fewer parameters and requires less computation during training and inference. Moreover, ConvGRU also tends to converge faster than ConvLSTM in our task. The reason is that the update gate in the ConvGRU allows the model to learn how much of the previous hidden state should be passed forward to the current hidden state. This reduces the risk of vanishing gradients and makes it easier to propagate gradients through time.

Third, we employ the power of transformers instead of the Global pooling layer or LSTM as used in existing works. The reasoning behind using a transformer is its ability to process input sequences of varying lengths while attending to specific parts with varying levels of detail, enabling it to capture complex temporal relationships between skeleton joints and make precise predictions.

The main contributions of this paper are summarized as follows:
\begin{itemize}
    \item A dense STGC-GRU model is proposed for end-to-end assessment of rehabilitation exercises;
    \item A ConvGRU layer is employed as an alternative to ConvLSTM to lower computation during training and inference;
    \item A transformer encoder architecture is proposed to overcome basic LSTM limitations;
    \item The proposed system offers clear guidance on which body parts or movements to focus on and enhance assessment quality, based on a self-attention mechanism;
    \item The efficiency of the proposed model is shown through extensive experimentation on two physical rehabilitation datasets, KIMORE and UI-PRMD.
    
\end{itemize}

The rest of the paper is structured as follows: In Section ~\ref{sec2}, related work on rehabilitation exercise assessment and the motivation for our proposal are discussed. Section ~\ref{sec3} provides a thorough explanation of the proposed system. Section ~\ref{sec4} outlines the experimental setup and results. Finally, concluding remarks and future perspectives are presented in Section ~\ref{sec5}.

\section{Related Works}\label{sec2}
Current methods of evaluating movement involve comparing a patient's exercise performance to that of healthy individuals. A recent study by  \cite{liao2020review} reviewed various computer-based techniques for evaluating patient rehabilitation exercises using motion tracking technology. Approaches evaluating patient rehabilitation exercises can be divided into three categories: (1) discrete movement score approaches (2) rule-based approaches (3) template-based approaches. In the following, we give a brief review of these different approaches.

\subsection{Discrete movement score approaches}
Using machine learning, these studies employ a discrete movement score in order to distinguish between two classes: correct and incorrect movement sequence classes. Generally, they output a binary class value for the given test patient sequences. Using such motion classification system to evaluate post-stroke rehabilitation, k-nearest neighbors \citep{zhang2011template}, Adaboost classifier \citep{taylor2010classifying}, random forest \citep{patel2010novel} or multi-layer perceptron neural networks \citep{jung2008feature} were used.

For home-based physiotherapy exercises assessment, Upper-Limb motor function impairment, Bayesian classifier and support vector machines (SVM)  were used \citep{ar2014computerized, otten2015framework}.
Deep and Convolutional Neural Networks \citep{um2018parkinson} were also used to diagnose Parkinson's disease using data from a wrist-worn wearable sensor. Despite their high accuracy, these methods cannot monitor changing movement quality or track improvement in patient performance during rehabilitation. Therefore, this category is not adequate for a robust and accurate  rehabilitation system. 

\subsection{Rule-based approaches}

Rule-based approaches for assessing rehabilitation exercises involve using predefined rules and criteria to evaluate and quantify a patient's performance. These approaches rely on clinical guidelines and best practices, tailoring the rules to the patient's specific needs and condition. Examples include the Functional Independence Measure (FIM) for neurological impairments and the Knee Injury and Osteoarthritis Outcome Score (KOOS) for knee disorders \citep{nolan2022post}. These approaches employ standard rules to assess patient movement, such as monitoring knee and ankle angles \citep{bo2011joint} or defining kinematic rules \citep{zhao2014realtime}. By providing a standardized and objective assessment, rule-based approaches ensure consistency and accuracy across evaluators and settings. However, they can be time-consuming and may not consider individual patient goals. Additionally, they may not be applicable to atypical cases or new conditions, limiting their generalizability. These approaches are particularly useful for simple exercises, but their effectiveness diminishes with exercise complexity. Moreover, they do not adapt well to novel exercises.

\subsection{Template-based approaches}
To avoid the need for rule-making and better reflect the patient's motor ability, these methods rely on a direct comparison between the patient and a template motion and employ distance function-based techniques. Distance metrics like Euclidean, Mahalanobis, and Hausdorff distances are used to measure similarity \citep{benetazzo2014low, houmanfar2014movement, huang2014using}. Generally in such solutions, Dynamic Time Warping (DTW) ensures sequence length invariance \citep{saraee2017exercisecheck}. Another group of researchers suggested probabilistic methods. They involve Gaussian mixture models to evaluate movement quality and detect deviations from ideal motions. The log-likelihood of individual sequences generated from a trained Gaussian mixture model is used for movement evaluation \citep{elkholy2019efficient}. Gaussian mixture models are also utilized to represent ideal movements in various contexts, such as detecting body part motion deviations \citep{gorer2017autonomous} and addressing low back pain rehabilitation \citep{devanne2017multi}. Discrete Hidden Markov Models (HMM) and Hidden Semi-Markov Models (HSMM) were proposed for segmenting and analyzing human motion data in physical therapy exercises \citep{wei2019towards, capecci2018hidden, osgouei2020rehabilitation}. Besides, \cite{williams2019assessment} used autoencoder neural networks to reduce high-dimensional motion trajectories to a low-dimensional space, followed by Gaussian mixture models for modeling movement density. Moreover, performance metrics based on the log-likelihood of Gaussian mixture models are introduced to encode low-dimensional data representations achieved with deep autoencoder networks \citep{liao2020deep}.

\subsection{Deep Learning based approaches}

Feature extraction from motion sequences in the context of exercise assessment has been approached through manual selection, traditional feature engineering algorithms such as manifold learning or PCA \citep{devanne2017multi, tao2016comparative, akremi2022spd}. While neural network architectures have been extensively explored for modeling human motion in other contexts like action recognition, only a few studies have focused on sequence motion assessment for patient rehabilitation exercises \citep{sun2022human}. Some researchers have proposed neural network architectures for encoding data features. For instance, \cite{vakanski2016mathematical} introduced an architecture consisting of an autoencoder subnet for dimensionality reduction and a mixture density network (MDN) to obtain probabilistic models of human motion. \cite{zhu2019deep} proposed a combined Dynamic Convolutional neural network (D-CNN) and State transition probability CNN (S-CNN) to address data alignment and capture discriminative exercise features. \cite{liao2020deep} presented a temporal-pyramid model that combines CNN and Recurrent Neural Networks architectures (RNN), incorporating spatial information from different body parts. Various methods using GCNs have also been proposed, leveraging the graph structure of human body skeleton data for action quality assessment and exercise evaluation \citep{song2020richly, zhang2020semantics, du2015hierarchical, li2018co}.
In the field of rehabilitation exercise assessment, GCNs have shown promise. \cite{deb2022graph} developed a spatio-temporal GCN for predicting continuous scores in exercise assessment. \cite{chowdhury2021assessment} used a GCN for spatial feature extraction and an LSTM network for temporal feature extraction from skeletal data to predict exercise quality. In the domain of action recognition, GCNs have been successfully employed, with architectures inspired by Spatio Temporal-GCN \citep{st-gcn, as-gcn, non-local-gcn}.
Inspired by the success of GCN-based methods and the potential for precise feedback on the human body skeleton, the proposed approach aims to respect the topological structure of skeleton data, and adopt a robust graph-based approach for exercise assessment. By incorporating graph-based techniques and transformers, this approach has the potential to provide accurate and visual feedback for evaluating rehabilitation exercises. 

\section{Proposed Approach}\label{sec3}
\subsection{Overview of the approach}\label{sec31}

\begin{figure*}[ht]
\centering
\includegraphics[width=12cm,height=9.5cm]{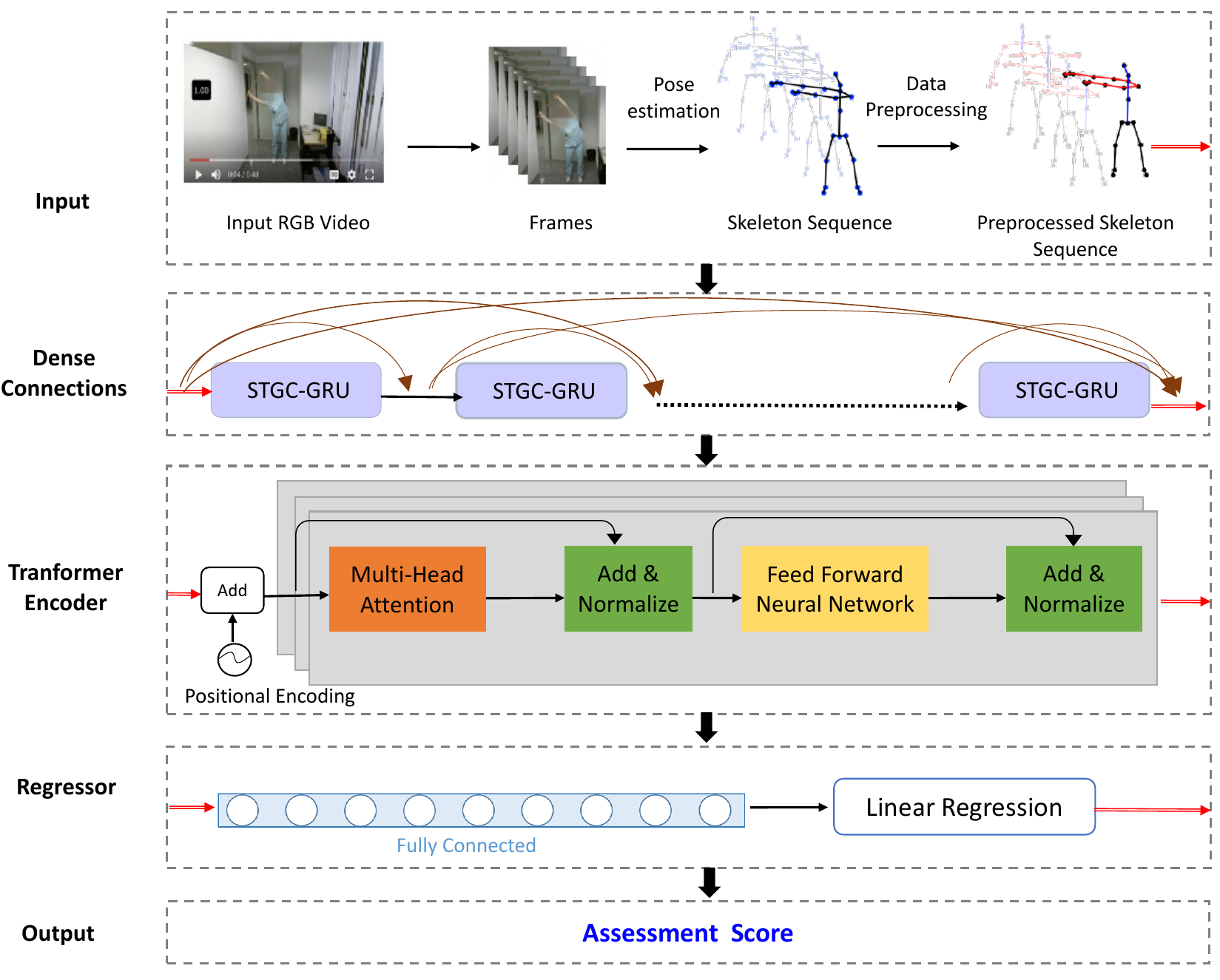}
\caption{Flowchart of the proposed approach.}
\label{fig:flowchart}
\end{figure*}

The flowchart of our proposed approach for exercise assessment is depicted in Figure ~\ref{fig:flowchart}.

The pipeline of the proposed framework is composed of four consecutive blocks.
It starts with the acquired input data containing RGBDs sequences with patients performing the needed evaluation exercises. The considered modality in our work is the skeleton data which can be given directly by the acquisition device (such as kinect) or deduced from RGB video using skeleton estimation approaches \citep{shotton2012efficient,pavllo20193d,bazarevsky2020blazepose}.
After the data preprocessing step on skeleton sequences, we propose to extract spatio-temporal features with a Dense Graph Convolutional connection layer. Inspired by Spatio-Temporal Graph Convolutional Networks structure (STGCN) by \cite{huang2020spatio}, we construct a network composed of multiple STGC-GRU blocks with direct connections from the output of each STGC-GRU block to all the output of the other blocks. Consequently, the output of the $i^{th}$ block receives the feature-maps of all preceding blocks. If we consider $ F_0, F_1, ..., F_{M-1}$ as the concatenation of the feature-maps produced in STGC-GRU blocks $0, 1, ..., M-1$, we have : 
\begin{equation}\label{Eq:1}
G_{N} = STGC-GRU([F_0, F_1, ..., F_{M-1}]) 
\end{equation}
The global representation extracted from STGC-GRU blocks, using Equation \ref{Eq:1}, is then used to feed the proposed transformer encoder. Finally, a fully connected layer is employed to predict the needed continuous assessment score. 

The network architecture is designed to take into account the sequential dependencies among the spatio-temporal features across frames/body movements by incorporating dense connections between STGC-GRU blocks. This enhances the propagation of spatio-temporal features and promotes feature reuse across various STGC-GRU blocks. Additionally, users can perform the same workout at varying speeds (slow or fast), causing differing spatiotemporal characteristics. To address this challenge, a transformer is employed to account for the varying spatiotemporal features of identical exercises. 

Besides, transformer models have been used for sequential data because they can learn those long-distance relationships but do not incorporate the topological structure of the human skeleton. Therefore, we propose to combine these two kinds of networks for physical rehabilitation. Our model takes advantage of an STGC-GRU architecture with a self-attention mechanism from the transformer encoder, to calculate the score of each pair of joints and updated the attributes of the current vertex.
The commonly used symbols and notations, in equations and figures of our paper, are summarized in the Table \ref{Tab:OverviewTerminology}.

\begin{table}[h!]
\centering
\begin{adjustbox}{max width=0.5\textwidth}
\begin{tabular}{c c } 
 \hline
Notation & Definition\\
 \hline
$\oplus$ & The concatenation operation\\
$\otimes$ & The convolution operation \\
$\odot$ & The element-wise product\\
$\phi$& The normalizing factor\\
$\times$ & The Hadamard product\\
$F$& The feature map\\
$V$& An RGBD video\\
$X$& A frame in the video sequence\\
$T$& The number of frames\\
$M$& The self attention map\\
$G$& The graph structure\\
$N_G$& The set of nodes of $G$\\
$E_G$& The set of edges of $G$\\
$A$& The adjacency graph matrix for $G$\\
$\tilde{A}$& The nomalized adjacency graph matrix\\
$D$& The degree matrix\\
$I$& The identity matrix\\
$P$& The processed video representation\\
$K_a$ & A kernel function\\
$W$& The learnable model parameters\\
$tanh$ & The hyperbolic tangent activation function\\
$\sigma$ & The sigmoid functions\\
$z_t$& The update gate in GRU bloc\\
$r_t$& The rest gate in GRU bloc\\
$o_t$ & The hidden state candidate in GRU bloc\\
$h_t$ & The hidden state output in GRU bloc\\
$Z$& The output tensor of STGC-GRU\\
$Q$& The query component in attention mechanism\\
$K$& The keys component in attention mechanism\\
$V$& The values  component in attention mechanism\\
$L$& The loss function\\
$y$& The true values \\
$\Tilde{y}$&  The predicted values \\
 \hline
\end{tabular}
\end{adjustbox}
\caption{Summary of commonly used notations.}
\label{Tab:OverviewTerminology}
\end{table}

\subsection{Problem Formulation}

An arbitrary exercise of rehabilitation is denoted by $V_i = \{X_{t=1...T}\}$, where $V_i$ refers respectively to the \textit{ith} RGBD video, $X_t$ is the $t^{ith}$ frame and $T$ is the number of frames. For each video, we associate a ground-truth performance score $y_i \in [min_{score},max_{score}]$ that represents the exercise quality.

We choose skeleton-based data encoding since it is more robust than RGB image-based modality to changes in body sizes, motion rates, camera perspectives, and interference backgrounds. Given a sequence of skeletons, we consider $N$  the number of joints representing each skeleton, where each joint has C-dimensional coordinates estimated by a pose estimation approach or encoded with the sensor that helped to capture the data. Dimension becomes for each video: $V_i \in \mathbb{V}^{T \times N \times C}$ and for each frame: $X_t \in \mathbb{R}^{N \times C}$.

For a given exercise, each skeleton motion plays an essential role depending on the performed exercise. Our goal is to predict  the score $\hat{y}_j$ to give the patient an idea about the quality of his performance. We also capture the role of all joints and give a feedback that assists the patient to improve the fluency of his exercise. Therefore, we consider a self-attention map $M_j \in \mathbb{R}^{T \times N \times N}$. The latter helps the patient to improve his performance by highlighting articulations, denoted by joints, where improvement is needed.

\subsection{Dense Spatio-Temporal Feature Extraction }
\begin{figure*}[ht]
\centering
\includegraphics[width=12cm,height=4cm]{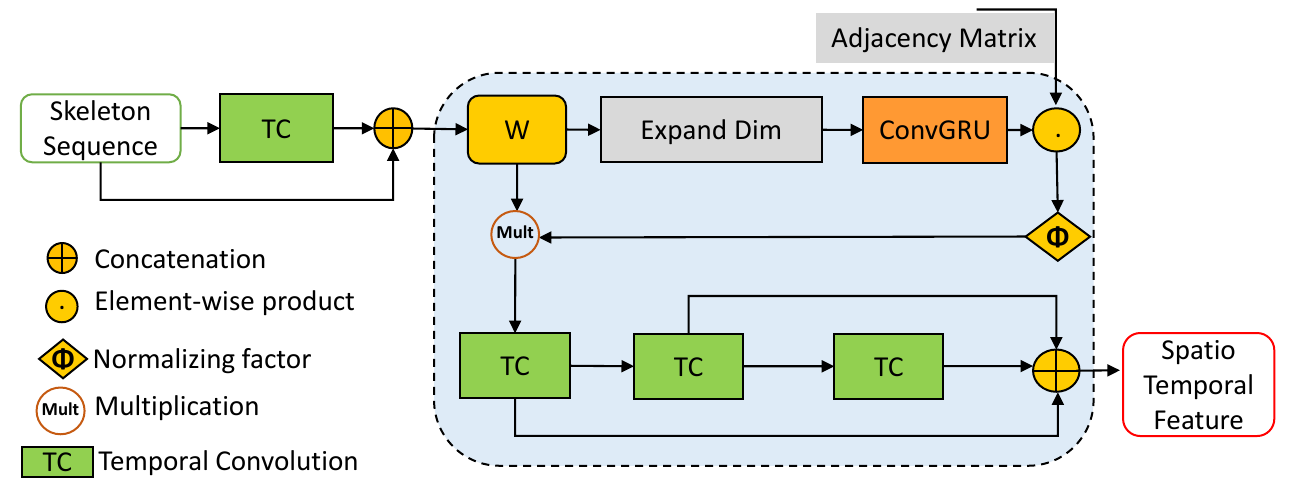}
\caption{STGC-GRU block details.}
\label{fig:stgcn}
\end{figure*}

 The skeleton can be viewed as a directed acyclic graph with a natural structure, using biomechanical dependencies between joints and body parts. Each joint is depicted as a node in the graph and connected to other joints via edges. Each of these joints has different features, such as the 3-dimensional coordinates and/or Euler angles. These values are given for each image (frame) belonging to a video sequence.
Recent studies demonstrate that representing human skeleton data as a graph is a natural choice to extract spatio-temporal features which characterize the best topological structure of the body joints connection.
Particularly, Graph Convolutional Networks (GCNs) have been used successfully in the field of human skeleton motion analysis as relational networks \citep{feng2022comparative}. Inspired by the recent Spatio-Temporal Graph Convolutional Networks STGCN by \cite{huang2020spatio}, we propose an extension of STGCN  for evaluating the effectiveness of physical therapy exercises. Our extended architecture utilizes graph convolutions with a dynamic adjacency matrix, building upon STGCN's original use for recognizing actions based on skeleton data. In this paper, we propose a Dense STGC-GRU block as illustrated in Figure ~\ref{fig:stgcn}.

Each frame of the input skeleton sequence is represented by its graph structure $G = (N_G,E_G,A)$, where $N_G=\{n_i\}_{i=1..25}$ is the set of nodes, $E_G$ is the set of edges and $A$ is the adjacency matrix of the graph.
 We formulate our adjacency matrix A\textsubscript{k} as follows:
 
\begin{equation}
A\textsubscript{k} = D\textsubscript{k}^{-1/2}.(\tilde{A\textsubscript{k}} + I).D\textsubscript{k}^{-1/2}
\label{eq:AdjMatrix}
\end{equation}

\begin{figure*}[ht]
\centering
\includegraphics[width=12cm,height=4cm]{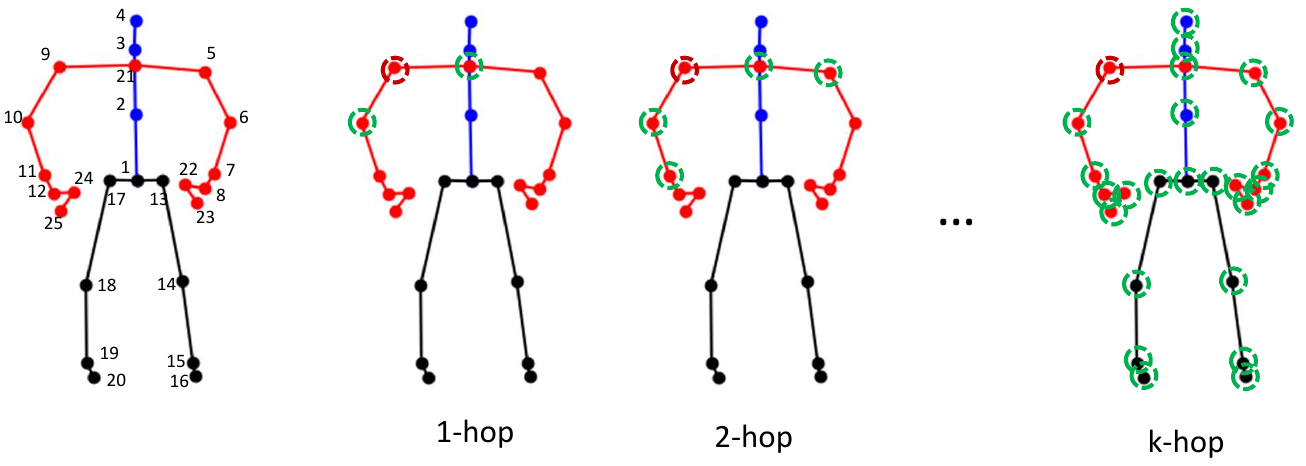}
\caption{First skeleton represents the 25 joints of a skeleton from KIMORE dataset. The other skeletons represent the nodes (in green) involved in the computation of the \textit{$1^{st}$}, \textit{$2^{nd}$} and \textit{$k^{iem}$} hop order regarding a certain joint (in dark red).}
\label{fig:hops}
\end{figure*}

 $\tilde{A\textsubscript{k}}$ is the adjacency matrix of our graph representing the connections between the skeleton nodes. An identity matrix $I$ is added to represent the self-connections of the nodes. ($\tilde{A\textsubscript{k}} +I$) is multiplied by $D\textsubscript{k}^{-1/2}$ (the inverse of the degree matrix of the graph) on both sides to normalize it.
We define the k-adjacency matrix $\tilde{A}_{(k)}$  as:
\begin{equation}
%[
(\tilde{A}_{(k)} )_{i,j} =
  \begin{cases}
    1       & \quad \text{if } d(n_i,n_j)=k,\\
    1  & \quad \text{if } i=j,\\
    0  & \quad otherwise.
  \end{cases}
%]
\label{eq:eq3}
\end{equation}
Where $d(n_i,n_j)$ gives the shortest distance in number of hops between node $n_i$ and node $n_j$.
$\tilde{A}_{(k)}$, in Equation \ref{eq:eq3}, is thus a generalization of
$\tilde{A}$
to further neighborhoods, with $\tilde{A}_{(1)} = \tilde{A}$ and $\tilde{A}_{(0)} =I$. Figure ~\ref{fig:hops} shows the adopted distance partitioning strategy for different k-hops.

First, we process the input sequence $V$ as $P=V \OPLUS (K_a \otimes V)$, where $\OPLUS$, $\otimes$, and $K_a$ denote respectively the concatenation operation, the temporal convolution operation, and the used kernel.

Second, a graph convolution is conducted as follows:

\begin{equation}\label{eq:4}
G(P)  =\sum_{k}^{\Gamma \textsubscript{a}}(P.A\textsubscript{k}).W\textsubscript{k},
\end{equation}
Where $\Gamma \textsubscript{a}$ is the kernel size on the spatial dimension, which also matches the number of adjacency matrices. The number of adjacency matrices depends on the used partitioning strategy, which we will explain below. $W\textsubscript{k}$ is a trainable weight matrix and is shared between all graphs to capture common properties. In Equation \ref{eq:4}, each kernel $P.A\textsubscript{k}$ computes the weighted average of a node's features with its neighboring nodes, which is multiplied by $(P.A\textsubscript{k}).W\textsubscript{k}$, a weight matrix. The features generated by all kernels are then summed to form a single feature vector per node. This operation helps extracting spatial features from the non-linear structure of the skeletal sequence.  It is inspired by the graph convolution from ST-GCN \citep{st-gcn}, which uses a similar Graph Convolution formulation to the one proposed by \cite{gcn}. 

Third, an improvement of STGCN block is proposed in this paper and consists of adding a Convolutional Gate Recurrent Unit (ConvGRU) layer. This layer helps to calculate a self-attention map which makes the adjacency matrix dynamic and is recomputed each time through the added layer.
GRU includes gates in one unit as LSTM does within a simpler structure. Thus, GRU is computationally cheaper.  This is very important in our application study and the benefit of using ConvGRU instead of ConvLSTM is studied in experimental results comparing the computation time and accuracy on used metrics.
ConvGRu combines CNN and GRU and thus has the advantage of maintaining the spatial structure of the skeletal input sequence and it is also more conductive for spatial-temporal features in time series. 

With an action sequence $\{x_{t=1...T}\}$ with $T$ frames, it performs the forward propagation as follows:
\begin{equation}  \label{eq:zt}
    z_t = \sigma(w_{zx}\otimes x_t + w_{zh}\otimes h_{t-1}+b_z) 
\end{equation}
\begin{equation} \label{eq:rt}  
    r_t = \sigma(w_{rx}\otimes x_t + w_{rh}\otimes h_{t-1}+b_r)
\end{equation} 
\begin{equation}\label{eq:ot}
     o_t = tanh(w_{ox}\otimes x_t + w_{oh}\otimes (r_t \times f_{t-1}) +b_o)
\end{equation}
\begin{equation}\label{eq:ht}
h_t = z_t\times x_t + (1-z_t)\times o_t
\end{equation}
We denote by $\otimes$ and $\times$ respectively convolution operation and Hadamard product. $tanh$ and $\sigma$ are tangent and Sigmoid functions. $w_x$, $w_h$ and $b$ are corresponding weights and biases. Equation \ref{eq:ht} represents the hidden state for each time index $t=1..T$ ($h_0$ is set to $0$) and it is considered as output and background information going in the network. Different gates in GRU are represented by $z_t$, $r_t$, and $o_t$ represented in Equations \ref{eq:zt}, \ref{eq:rt} and \ref{eq:ot}.

Afterward, we proceed to inject an adjacency matrix, as derived from Equation ~\ref{eq:AdjMatrix}, into the ConvGRU output through elementwise multiplication, followed by the application of a normalization factor.

Finally, in order to extract different levels of temporal  features, three Temporal Convolutional layers are performed consecutively and their respective output is concatenated to give the needed spatio-temporal features. We use several consecutive STGC-GRU blocks with the same structure to enable a better capture of different representation levels of nodes in the network. Besides, we propose to consider a dense network that uses shortcut connections. These dense connection operations help to: (i) learn the mapping between the information of previous and next feature levels (ii) enable the reuse of
contextual information at different scales so that the network can capture abundant spatial-temporal contextual information (iii) capture richer dependencies among joints and retain more structural information of human pose, leading to satisfactory performance.

\subsection{Position Encoding}
The tensor obtained by STGC-GRU blocks does not contain the order of joint tokens, and the identity of joints cannot be distinguished, making the self-attention unable to capture the sequential characteristic of movements, which will reduce the performance of movement scoring. The experimental results confirm our hypothesis in Table ~\ref{Tab:positionencoding}.

To solve this issue, \cite{vaswani2017attention} suggest using position encoding to label each joint and applying sine and cosine functions with varying frequencies as the encoding functions :

\begin{equation}\label{eq:Ecod1}
    PE(x,2i) = sin(x/10000^{(2i/D)})
\end{equation}
\begin{equation}\label{eq:Ecod2}
    PE(x,2i+1) = cos(x/10000^{(2i/D)})
\end{equation}

Where $i$ ranges from $0$ to $\frac{d}{2}$ and $d$ represents the input dimension. This sinusoidal position encoding enables the transformer to model the position of a joint token and the distance between each pair of joint tokens.

\subsection{Transformer encoder block for variable-length and smoothness}

Rehabilitation exercises data show notable variability within variable-length data, in contrast to related issues with sequential data. One significant factor is that the exercise participants are generally diverse individuals, ranging from experienced therapists to patients with various illnesses and disabilities. Additionally, the number of repetitions required for rehabilitation exercises may vary depending on the therapist's prescription. As a result, different users assign the same workout with the same number of repetitions with varied lengths of time to complete. A recent work of \cite{yan2018spatial} extracted spatio-temporal features by employing a global pooling layer that comes before the FC layers, ignoring the spatio-temporal characteristics' underlying sequential relationships between frames/body movements. As a result, users who execute the identical activity quickly or slowly provide various spatio-temporal information. To overcome these limitations, \cite{deb2022graph} use LSTM architecture to capture sequential dependencies that exist in spatio-temporal features, to extract discriminative features that have accumulated over time. In this work, we employ a transformer architecture instead of LSTM.

Transformers handle variable-length input sequences because they use self-attention mechanisms, which allow the model to weigh the importance of different parts of the input sequence without requiring a fixed-length context. This means that the model can adapt to the specific length and structure of the input, rather than being limited by a fixed-length context window or requiring the input to be padded to a fixed length. This allows for more flexible and efficient processing of input sequences of varying lengths. Moreover, transformers can learn the relationships between each element of a sequence, thanks to their self-attention ability. It addresses the issue that LSTM and RNN networks struggle to accurately simulate long-term sequences by handling very long sequences. Furthermore, our model uses a multi-headed self-attention mechanism instead of traditional LSTM or RNN networks. Unlike token-by-token processing in these networks, the self-attention mechanism allows parallel processing of sentences. This enables efficient calculation of joint correlations in multiple consecutive frames, making self-attention a suitable choice for modeling skeleton data. The transformer blocks in our proposed network follow the scaled dot-product attention as proposed by \cite{vaswani2017attention}. Our transformer encoder takes the output tensor of STGC-GRU blocks $Z\in\mathbb{R}^{B,F,T}$, where $B$ indicates the batch size, $F$ is the number of the sequences, and $T$ denotes the sequence size. To start, we use positional encoding to assign a vector to each joint token. Then, we use three learnable matrices $W_q$, $W_k$, and $W_v$ to transform the joint data, $Z$, into separate spaces. These matrices typically have dimensions $\mathbb{R}^{B,F,dim}$, where $dim$ is a hyperparameter. Following this, we compute the attention for the query, key, and value matrices, $Q$, $K$, and $V$, respectively, using the following equations in each head:

\begin{equation}\label{eq:QKV}
    Q,K,V = ZW_q, ZW_k, ZW_v
\end{equation}

\begin{equation}\label{eq:Atj}
   A_{t,j} = Q_tK_{j}^T
\end{equation}

\begin{equation}\label{eq:A(QKV)}
    Attention(Q,K,V) = Softmax(\frac{A}{\sqrt{dim}})V
\end{equation}

In Equation \ref{eq:Atj}, $Q_t$ represents the query vector for the \textit{$t^{th}$} joint token and $j$ represents the joint token that the \textit{$t^{th}$} joint token attends to. $K_j$ is the key vector representation for the \textit{$j^{th}$} joint token. The softmax operation is applied along the last dimension. The ability of self-attention is improved through multi-head self-attention, which uses multiple groups of $W_q, W_k, W_v$ instead of just one group. Its formulas are as follows:

\begin{equation}\label{eq:QKV2}
    Q^{(h)},K^{(h)},V^{(h)} = ZW_q^{(h)}, ZW_k^{(h)}, ZW_v^{(h)}
\end{equation}

\begin{equation}\label{eq:head}
    head^{(h)} = Attn( Q^{(h)},K^{(h)},V^{(h)})
\end{equation}

\begin{equation}\label{eq:Multihead}
   MultiHead(H) = [head^{(1)}; ...; head^{(n)}]W_O
\end{equation}

Where $n$, in Equation \ref{eq:Multihead}, refers to the number of heads and $h$, in Equations \ref{eq:QKV2} and \ref{eq:head}, represents the head index. The concatenation in the last dimension is represented as $[head^{(1)}; ...; head^{(n)}]$. The learnable parameter $W_O$ has size $\mathbb{R}^{d \times d}$, where $d = dim \times n$.

The multi-headed self-attention mechanism maps a query to a series of key and value pairs, allowing it to model the relationship between input tokens after positional encoding. It considers the influence of node $n_i$ on other nodes and the impact of all other nodes on node $n_i$ when computing self-attention. The multi-head attention output is further processed through a basic feed-forward neural network. For faster training, layer normalization is employed instead of the commonly used batch normalization in standard feed-forward neural networks. Moreover, residual connections are used in our transformer encoder block. For instance, smoothness is a crucial factor in determining how well an exercise is scored. To determine how smooth a movement is, we must look at the temporal characteristics (velocity, acceleration) of the total conjugative time frames. A pooling or an LSTM layer may fail to capture the complete smoothness information, which is crucial in determining the correctness score. By using a transformer encoder block, smoothness information from past to future movements can be captured in parallel, yielding improved results as the dependencies are better understood.

\subsection{Proposed Losses}
To solve the regression problem, our proposed network is trained with various regression losses, including Mean Square Error (MSE), Huber Loss, and Log-Cosh Loss. During inference, a test sequence of skeleton data is processed to compute a continuous assessment score using these losses. The following provides a description of each one.\\

\textit{Mean Square Error Loss:} \\
Mean square error (MSE) is the most widely used regression loss function in machine learning. It measures the sum of the squared difference between predicted and target values. A lower MSE indicates a better-performing regression model. MSE is calculated as the average sum of squared differences between the actual value and the value estimated by the regression model.

\begin{equation}\label{eq:MSE}
    MSE= \frac{\sum_{i=0}^n (y_i-\hat{y}_i)}{n}
\end{equation}

Where $y_i$ and $\hat{y}_i$ denote the actual value and the predicted value respectively and $n$ represents the number of samples. \\

\textit{Huber Loss:} \\
Huber loss is a regression loss function that combines the benefits of $l_2$ and $l_1$ penalties. It is less affected by outliers in data compared to other losses, and transitions from absolute error to quadratic error based on a hyperparameter, $\delta$. The smaller the error, the more it becomes quadratic. Huber loss approaches MSE, presented in Equation \ref{eq:MSE}, as $\delta\rightarrow0$ and mean absolute error (MAE) as $\delta\rightarrow\infty$. The formula for Huber loss is as follows:

\begin{equation}\label{eq:Loss1}
    L(y_i-\hat{y}_i) = \begin{cases}
        \frac{1}{2}(y_i-\hat{y}_i)^2 & ; \arrowvert(y_i-\hat{y}_i)\arrowvert\le \delta\\
        \delta \arrowvert y_i-\hat{y}_i \arrowvert -\frac{\delta}{2}  & ; otherwise
     \end{cases}
\end{equation}
The selection of $\delta$ is crucial as it determines what is considered an outlier. Huber loss is advantageous in such scenarios as it smoothly bends around the minimum, reducing the gradient. Additionally, it is more resilient to outliers than MSE.\\

\textit{Log-Cosh Loss:} \\
Log-cosh is another function used in regression tasks that is smoother than MSE. It is the logarithm of the hyperbolic cosine of the prediction error which can be defined by the formula below : 
\begin{equation}\label{Eq:Loss2}
L(y_i-y_{i}^p) = \sum_{i=0}^n log(cosh(y_i-y_{i}^p))
\end{equation}

The Log-Cosh Loss operates similarly to MSE but is less impacted by occasional large errors. Like Equation \ref{eq:Loss1}, it has all benefits of Huber loss but also has the advantage of being twice differentiable everywhere, unlike Huber loss. \\

\subsection{Network architecture}
In this section, we present the detailed layers description of our proposed model (D-STGCNT). Our STGC-GRU block constitutes a temporal convolution with 64 kernels of size (9,1), followed by the ReLU activation layer. Then, the output of the temporal convolution is concatenated with the first input sequence to produce a tensor $Z$. Subsequently, a graph Conv-GRU with 64 and 25 kernels of size (1,1) and a GRU layer is utilized on $Z$ and the \textit{$k^{th}$} hop adjacency matrix to capture spatial characteristics from the topological layout of human skeletons. This is followed by three temporal convolutional layers with equal padding and kernels of size (9,1), (15,1), and (20,1), respectively, with 16 filters for each layer. The output of our extended STGC-GRU block is the concatenation of the three temporal convolution outputs. The purpose of concatenating is to identify movement patterns at varying levels of abstraction. Our work uses dense connections in multiple STGC-GRU blocks to extract more intricate features, facilitating spatiotemporal feature propagation across layers, and promoting feature reuse across various STGC-GRU blocks. 
The output of the latter is processed by a positional encoding module  to incorporate the order of sequences. Then, the two terms are added together as follows: $Output= Output^{STGC-GRU} + PosEncoding(Output^{STGC-GRU})$. Finally, numerous transformer encoder blocks are employed as presented in Figure ~\ref{fig:flowchart}, where each block constitutes a layer normalization with $\epsilon=1e-$ instead of using batch normalization, to stabilize the network which results in substantially reducing the training time necessary. Then, we use a multi-head attention layer where $head-size = 128$ and $num-heads =6$. We add next a dropout layer (dropout = 0.1) in order to avoid over-fitting. Two Feed forward layers of Conv1D with 80, and 128 kernels are employed instead of the Dense layers which are used on traditional transformers. The concept behind using Conv1D is to enhance the representation of attention outputs through projection. Additionally, residual connections are employed between layers to facilitate network training by facilitating gradient flow. 

By stacking transformer encoders $N$ times, we increase information encoding. Every layer has the chance to learn distinct attention representations, increasing the power of the attention network. The result of the stacked transformer encoders is then processed by a linear activation fully connected layer.

\section{Experimentation and Results}\label{sec4}

In this section, we conducted extensive comparative experiments to evaluate the performance of our model (D-STGCNT). First, we describe rehabilitation exercise datasets and metrics used for evaluation. Then implementation details are introduced. In the following, we conducted extensive ablation studies to verify the contribution of the individual components of our D-STGCNT. Finally, we quantitatively compare our proposed approach with several state-of-the-art methods.

\subsection{Evaluation Process}
 In the following, we present datasets used to assess the effectiveness of our proposed model. Then, we introduce the evaluation metrics employed to measure model performance. The implementation details are also presented in the paper, including the programming language used, any relevant libraries, and the pre-processing steps applied to the data.

\subsubsection{Dataset}
\label{sec5}
Extensive experiments are conducted on rehabilitation exercises from two datasets (see Table ~\ref{Tab:Datasets}).

\begin{itemize}
    \item KIMORE \citep{capecci2019kimore}: This dataset includes RGBD videos and score annotations for five exercises, divided into two groups: control (expert and non-expert) and pain/postural disorder (Parkinson, back-pain, stroke). The control group has 44 healthy subjects, with 12 being physiotherapists and experts in rehabilitation and 32 being non-expert. The pain/postural disorder group consists of 34 subjects with chronic motor disabilities.

    \item UI-PRMD \citep{vakanski2018data}: This dataset is publicly available and contains movements of common exercises performed by patients in physical rehab programs. Ten healthy individuals performed 10 repetitions of various physical therapy movements, captured using a Vicon optical tracker and a Microsoft Kinect sensor. The data includes full-body joint positions and angles, and its purpose is to serve as a foundation for mathematical modeling of therapy movements and establishing performance metrics to evaluate patients' consistency in executing rehabilitation exercises.

\end{itemize}

\begin{table*}[h!]
\centering
\begin{adjustbox}{max width=1\textwidth}
\begin{tabular}{c c c c c c c} 
\hline
Features &  Sensor &  Depth Imaging System & \# of Subjects & \# of Exercises & Range of Quality Scores \\ 
\hline
UI-PRMD \cite{vakanski2018data} & Vicon and Kinect v2 & skeleton  & 10 & 10 & 0-1 \\
KIMORE \cite{capecci2019kimore}  & Kinect v2 & RGB-D and skeleton  &  78 & 5 &  0-50 \\
\hline

\end{tabular}
\end{adjustbox}
\caption{UI-PRMD and KIMORE datasets description.}
\label{Tab:Datasets}
\end{table*}

\subsubsection{Evaluation metrics}

To evaluate and compare our approach with state-of-the-art methods,  we use the metrics used by \cite{liao2020deep} and \cite{deb2022graph}. If $y$ is our target, $\hat y$ is our prediction, $n$ the number of observations and $e= \hat{y} - y $ is the error
, then used metrics can be defined as follows:

\begin{itemize}
    \item Mean Absolute Deviation (MAD): average of the absolute deviation between ground truth values and predicted values: $ MAD=median(e - median(e)).$

    \item Mean Squared Error (MSE) and Root Mean Squared Error (RMSE) which are defined as follows:

     $MSE = \frac{1}{n} \sum_{i=1}^n e_i^2 $ and  $RMSE = \sqrt{MSE} $.
     \\These are the most common regression metrics. They are very sensitive to outliers and penalize large errors more heavily than small ones. 
    \item Mean Absolute Percentage Error (MAPE) measures the percentage error of the forecast in relation to the actual values: 
    $MAPE = \frac{100\%}{n} \sum_{i=1}^n \displaystyle\left\lvert \frac {y_i – \hat{y}_i}{y_i } \right\rvert$
\end{itemize}

The proposed approach is designed for patients seeking a quick and accurate indication of the quality of the rehabilitation exercises they are performing, therefore we have also calculated the response time of our proposed approach both in the train and test phases while being watchful for any possibility of optimization. Furthermore, we also evaluated visually and by interpretation the quality of the feedback given by our model. In all our experiments, we follow the evaluation protocol defined by \cite{deb2022graph} for the division of the datasets into train and test parts.

\subsubsection{Implementation details}
The proposed D-STGCNT model has been implemented with python 3.6 using Tensorflow 2.x framework. We used a PC with Intel® Xeon® Silver 4215R CPU, with 32GB of RAM and a GeForce GTX 3080 Ti 16GB RAM graphics card. Our D-STGCNT is trained using Adam optimizer for 1500 epochs with batch sizes 10, and 3 for KIMORE and UI-PRMD datasets respectively. The learning rate is set to $1e-4$. We select the best model to assess the model's effectiveness on the test set in accordance with the validation set. To objectively assess the performance of our model in comparison to recent works, we additionally give the 10-run result, as like \cite{liao2020deep}, \cite{gcn}, and \cite{deb2022graph}. Ten times our model was performed on training and testing. To ensure the accuracy of our results, we save the performance measures (MAD, RMSE, and MAPE) from each run before averaging them.

\subsection{Ablation study}

\subsubsection{Effect of positional encoding }

We investigate the effect of the positional encoding as shown in Tab. ~\ref{Tab:positionencoding}. Results show that the performance of our model D-STGCNT without positional encoding is lower whit higher values of MAD, RMSE, MSE, and MAPE, and by Utilizing positional encoding, we enhance the performance significantly. This can be explained by the fact that different spatiotemporal joints play unique roles in action, and effectively utilizing this sequential information leads to significant improvement.

\begin{table}[h!]
\begin{adjustbox}{max width=0.5\textwidth}
\centering
\begin{tabular}{c c c c c} 
 \hline
Our model & MAD  & RMSE & MSE & MAPE\\
 \hline
Without positional encoding & 0.721 & 1.902  & 3.619 &  1.904 \\ 
With positional encoding & \textbf{0.399}  & \textbf{0.735}  & \textbf{0.540} &  \textbf{1.217} \\ 
 \hline
\end{tabular}\end{adjustbox}
\caption{Performance of our proposed model with and without positional encoding on Ex5 of KIMORE dataset.}
\label{Tab:positionencoding}

\end{table}

\subsubsection{Effect of the $k^{th}$ hop adjacency matrix}

In order to validate the effectiveness of the number of hops in our D-STGCNT model, we respectively set the number of hops to be {1,2}. 

Table ~\ref{Tab:kthop} presents the results of D-STGCNT with different hops. We observe that  combining (concatenating) the multiple hop adjacency matrix from different perspectives leads to better performances using the proposed metrics. Indeed, the concatenation represents long-range structural relations which leads to improving the movement assessment performance significantly.

\begin{table}[h!]
\centering
\begin{adjustbox}{max width=0.5\textwidth}
\begin{tabular}{c c c c c} 
 \hline
Our model & MAD  & RMSE & MSE & MAPE\\
 \hline
First hope only & 0.647 & 1.338  & 1.792 &  1.868 \\ 
Second hope only & 0.789 & 1.548  & 2.397 &  2.267 \\ 
Concatenate(First,Second) hope & \textbf{0.399}  & \textbf{0.735}  & \textbf{0.540} &  \textbf{1.217} \\ 
 \hline
\end{tabular}
\end{adjustbox}
\caption{Performance of our proposed model with different k-hops on Ex5 of KIMORE dataset.}
\label{Tab:kthop}
\end{table}

\subsubsection{Effect of regression losses}

The choice of the regression loss function in the training of our model can have a significant impact on its performance. Regression loss functions are used to measure the difference between the predicted output and the true output and are used to update the model's parameters during training. We evaluate
our model using three different regression losses MSE, Log-Cosh, and Huber loss. Results in  Table \ref{Tab:losses} show that Huber loss gives better results for predicting assessment scores. One of the main advantages of using Huber loss for training our model is that it can help to improve the robustness of the model. Another advantage of Huber loss is that it can provide a balance between MSE and MAE. MSE is sensitive to outliers and it penalizes large errors more heavily than smaller errors, while MAE is less sensitive to outliers and it gives equal weight to all errors. Huber loss is a combination of both, and it can provide the best solution, depending on the value of the Huber delta parameter.

Finally, Huber loss can also improve the stability of the optimization process during training. It is less sensitive to outliers, which can make the optimization process more stable and less likely to be affected by extreme values in the data. Varying the $\delta$ parameter in Huber loss can improve regression results by affecting the balance between mean squared error (MSE) and mean absolute error (MAE). A smaller delta value would make the loss closer to MSE, which is more sensitive to large errors and more appropriate for datasets with normally distributed errors. A larger delta value would make the loss closer to MAE, which is more robust to outliers and more appropriate for datasets with heavy-tailed errors. By tuning the delta value, we observe in Table ~\ref{Tab:losses} that with $\delta = 0.1$, we obtain the best results.

\begin{table}[h!]
\centering
\begin{adjustbox}{max width=0.5\textwidth}
\begin{tabular}{c c c c c} 
 \hline
Our model & MAD  & RMSE & MSE & MAPE\\
 \hline
With MSE loss & 0.623 & 1.291  & 1.668 &  1.784 \\ 
With Log-Cosh loss & 0.786 & 1.803  & 3.251 &  2.410 \\ 
With Huber loss ($\delta=1$) & 0.848  & 2.036  & 4.147 &  2.63 \\ 
With Huber loss ($\delta=0.1$) & \textbf{0.399}  & \textbf{0.735}  & \textbf{0.540} &  \textbf{1.217} \\ 
With Huber loss ($\delta=0.05$) & 0.522  & 1.023  & 1.046 &  1.635 \\ 
 \hline
\end{tabular}
\end{adjustbox}
\caption{Performance of our proposed model with different regression losses on Ex5 of KIMORE dataset.}
\label{Tab:losses}
\end{table}

\subsubsection{Effect of transformer vs LSTM}
The effect of utilizing a transformer architecture compared to LSTM in the context of online processing for patient's physical rehabilitation assessment is worth considering. Transformers offer the advantage of parallel processing, enabling more efficient online analysis of patients' physical rehabilitation. Transformers operate on self-attention mechanisms, allowing each element in the sequence to attend to all other elements simultaneously. These advantages enable faster and more efficiency in terms of performance and computations (5 times faster) than LSTM as shown in Table \ref{Tab:transformer}. This can be particularly beneficial in real-time patient assessment scenarios.

\begin{table}[h!]

\begin{adjustbox}{max width=0.5\textwidth}
\centering
\begin{tabular}{c c c c c c} 
 \hline
Our model & MAD  & RMSE & MSE & MAPE & Execution Time\\
 \hline
Our approach with  LSTM& 0.601 & 1.122  & 1.519 &  1.574 & 10.98 seconds\\ 
Our approach with Transformer & \textbf{0.399}  & \textbf{0.735}  & \textbf{0.540} &  \textbf{1.217}&\textbf{2.11} seconds\\ 
 \hline
\end{tabular}\end{adjustbox}
\caption{ Effect of using transformer and LSTM components on our model performance using test data from Ex5 of KIMORE dataset.}
\label{Tab:transformer}
\end{table}

\subsection{Comparison with state of the art}
To evaluate the effectiveness of the proposed model, a quantitative assessment was conducted by comparing it with several existing state-of-the-art approaches using identical datasets. The comparison was performed using the optimal parameters selected based on the results obtained from the ablation study. Our model incorporated positional encoding within the transformer architecture, employed concatenation of the first and second hop adjacency matrix, and utilized Huber loss (with $\delta =0.1$) as the regression loss function. Furthermore, the computational time of the proposed model was measured and compared against that of alternative methods, shedding light on the efficiency of the proposed approach.

\subsubsection{Quantitative comparison}

In Table ~\ref{Tab:MetricsComparison} and ~\ref{Tab:MetricsComparisonUI}, we present our results on MAD, RMSE and MAPE performances and those of the state of the art on the KIMORE and UI-PRMD datasets respectively. First, we report results for each of the five exercises included in the KIMORE dataset. Then, 
we report results using computed on the ten exercises of UI-PRMDE dataset.

We would like to point out that, in the Table ~\ref{Tab:MetricsComparisonUI}, comparisons are conducted on Kinect V2 joint position data instead of Vicon angles data as reported by \cite{liao2020deep}.
Our approach, which combines Dense STGCN, ConvGRU, and Transformer architectures, demonstrates superior performance in terms of MAD, RMSE, and MAPE. When compared to the approach proposed by \cite{deb2022graph}, which uses GCNs followed by an LSTM, our approach outperforms in terms of accuracy and precision. This improvement can be attributed  to the combined use of Dense STGCN and ConvGRU components, which effectively capture spatial and temporal features, and the Transformer architecture, which efficiently handles long-range dependencies and enables parallel processing. In comparison to \cite{song2020richly}, who use a multi-stream Graph Convolutional Network, our approach exhibits superior performance.  The integration of Dense STGCN, ConvGRU, and transformer architectures in our approach allows for a more comprehensive analysis of spatial and temporal features, resulting in enhanced accuracy and lower error metrics.

\begin{table*}[h!]
\centering
\begin{adjustbox}{max width=1\textwidth}
\begin{tabular}{c c c c c c c c c c} 
 \hline
Metric & Ex  & Ours&  \cite{deb2022graph}  &  \cite{song2020richly} & \cite{zhang2020semantics} & \cite{liao2020deep}& \cite{yan2018spatial}& \cite{li2018co}& \cite{du2015hierarchical}\\
 \hline
& Ex1 & \textbf{0.641}  & 0.799 & 0.977 & 1.75 7& 1.141 & 0.889 & 1.378 & 1.271\\ 
& Ex2 & \textbf{0.753}  & 0.774 & 1.282 & 3.139 & 1.528 & 2.096 & 1.877 & 2.199\\ 
MAD & Ex3 & \textbf{0.210}  & 0.369 &  1.105 & 1.737 & 0.845 & 0.604 & 1.452 & 1.123\\ 
& Ex4 & \textbf{0.206}  & 0.347 & 0.715 & 1.202 & 0.468 & 0.842 & 0.675 & 0.880\\ 
& Ex5 & \textbf{0.399}  & 0.621 & 1.536 & 1.853 & 0.847 & 1.218 & 1.662 & 1.864\\ 
 \hline
& Ex1 & 2.020 & 2.024 & 2.165 & 2.916 & 2.534 & \textbf{2.017} & 2.344 & 2.440\\ 
& Ex2 & \textbf{1.468}  & 2.120 & 3.345 & 4.140 & 3.738 & 3.262 & 2.823 & 4.297\\
RMSE &Ex3 & \textbf{0.487} & 0.556 & 1.929 & 2.615 & 1.561 & 0.799 & 2.004 & 1.925\\
& Ex4 & \textbf{0.527}  & 0.644 & 2.018 & 1.836 & 0.792 & 1.331 & 1.078 & 1.676\\ 
& Ex5 & \textbf{0.735}  &  1.181& 3.198 & 2.916 & 1.914 & 1.951 & 2.575 & 3.158\\ 
 \hline
& Ex1 & \textbf{1.623}  & 1.926 & 2.605 & 5.054 & 2.589 & 2.339 & 3.491 & 3.228\\ 
& Ex2 & \textbf{0.974}  & 1.272 & 3.296 & 10.436 & 3.976 & 6.136 & 5.298 & 6.001\\
MAPE &Ex3 & \textbf{0.613} & 0.728 & 2.968 & 5.774 & 2.023 & 1.727 & 4.188 & 3.421\\
& Ex4 & \textbf{0.541}  & 0.824 & 2.152 & 3.901 & 2.333 & 2.325 & 1.976 & 2.584\\ 
& Ex5 & \textbf{1.217}  & 1.591 & 4.959 & 6.531 & 2.312 & 3.802 & 5.752 & 5.620\\ 
 \hline
\end{tabular}}
\end{adjustbox}
\caption{
Results of our method in comparison with other state-of-the-art approaches on the KIMORE dataset.}
\label{Tab:MetricsComparison}
\end{table*}

\begin{table*}[h!]
\centering
\begin{adjustbox}{max width=1.1\textwidth}
\begin{tabular}{c |c c| c c| c c} 
 \hline
Metrics & \multicolumn{2}{|c|}{\textbf{MAD}} & \multicolumn{2}{c|}{\textbf{RMSE}} & \multicolumn{2}{c}{\textbf{MAPE}} \\ 
 \hline
Ex  & Ours& \cite{deb2022graph}  & Ours& \cite{deb2022graph} & Ours& \cite{deb2022graph}\\
 \hline
Ex1 & \textbf{0.011}  & 0.012 & \textbf{0.019} & 0.020 & \textbf{1.289} & 1.337  \\ 
Ex2 & \textbf{0.009}  & 0.011 & \textbf{0.014} & 0.016 & \textbf{1.105} & 1.244 \\ 
Ex3 & \textbf{0.013}  & 0.015 & \textbf{0.020} & 0.024 & \textbf{1.592} & 1.758  \\ 
Ex4 & \textbf{0.009}  & 0.010 & \textbf{0.011} & 0.015 & \textbf{0.984} & 1.090  \\
Ex5 & \textbf{0.009}  & 0.010 & \textbf{0.013} & 0.014 & \textbf{1.032} & 1.176 \\

Ex6 & \textbf{0.013} & 0.017 & \textbf{0.020}  &  0.025 & \textbf{1.476}  & 1.994 \\ 
Ex7 & \textbf{0.022}  & 0.023 & \textbf{0.034}  & 0.036 & \textbf{2.697}  & 2.980  \\

Ex8 & \textbf{0.020} & 0.024 & \textbf{0.032}  & 0.034 & \textbf{2.362}  & 2.815 \\

Ex9 & \textbf{0.013}  &  0.017 & \textbf{0.019}  & 0.022 & \textbf{1.455}  & 1.873 \\ 
Ex10 & \textbf{0.014}  & 0.025& \textbf{0.023}  & 0.033 &\textbf{1.619}  & 2.900\\
 \hline
\end{tabular}}
\end{adjustbox}
\caption{Results of our method in comparison with other state-of-the-art approaches on the UI-PRMD dataset.}
\label{Tab:MetricsComparisonUI}
\end{table*}

\subsubsection{Computational time}

Our proposed model is tested, in terms of computational time and accuracy, on the Ex5 of the KIMORE dataset in comparison with \cite{deb2022graph}. Results in Table ~\ref{Tab:time} show that it can provide real-time performance with a good score in terms of MAD, RMSE and MAPE. The computational cost was measured using a GeForce GTX 3080 Ti with 16GB of RAM. Indeed, our extended architecture uses ConvGRU layers which are generally considered to be faster.

ConvGRU offers faster processing due to fewer parameters, making it more memory-efficient and quicker to train. ConvGRU's efficiency in terms of speed is crucial for real-time analysis in patient rehabilitation. Its memory efficiency is advantageous in scenarios with limited resources, ensuring smoother execution and reducing the risk of bottlenecks. While ConvLSTM may excel in precise long-term modeling, ConvGRU, used in our experiments, showed sufficient performances outperforming state of the arts methods for tracking and analyzing patient movements during rehabilitation. Additionally, ConvGRU integrates seamlessly with the Jetson Nano platform and a camera, enabling real-time analysis and immediate feedback for patient rehabilitation.

Moreover, our model employs transformer encoders which are faster than LSTMs because they can process the entire 3D skeletons input sequence in parallel, and use an attention mechanism to selectively focus on relevant parts of the input.

\begin{table}[h!]
\centering
\begin{adjustbox}{max width=0.5\textwidth}
\begin{tabular}{c c c c c c c c} 
 \hline
&Phase & Number of Videos  &  Execution Time&MAD &RMSE &MAPE \\
 \hline
\cite{deb2022graph}&Train & 373 &  57 hours& - & - & - \\ 
Our model & Train & 373 & 25 min& - & - & - \\ 
 \hline
\cite{deb2022graph}&Test & 100 & 13.87 seconds& 0.631& 1.185& 1.602 \\ 
Our model & Test & 100  & 2.11 seconds& 0.404 & 0.739 &1.220\\ 
 \hline
\end{tabular}
\end{adjustbox}
\caption{Computational time for Ex5 of KIMORE dataset for 1500 epochs.}
\label{Tab:time}
\end{table}

\begin{figure*}[ht]
\centering
\includegraphics[width=12cm,height=3cm]{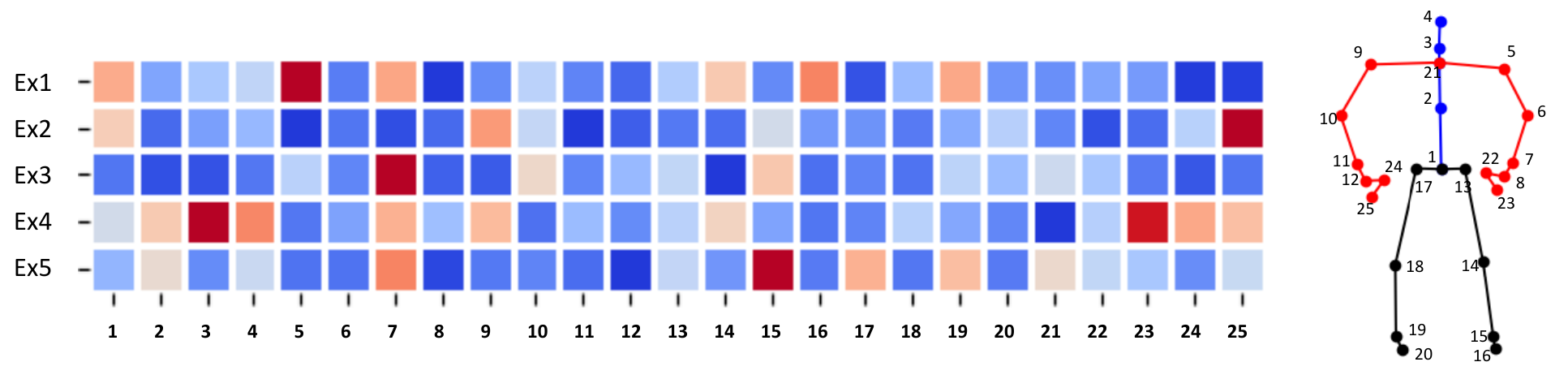}
\caption{An illustration of the attention value calculated by our approach that shows the involvement of the joints depending on the corresponding activities. On the left, we can see the calculated joints importance throw 5 exercises from the KIMORE dataset (lifting arms, arms extension, trunk rotation, pelvis rotation, squatting), and on the right, we illustrate the  25 joints of the skeleton human body as represented in KIMORE dataset.}
 \label{fig:ExercisesJointRoleVisu}
\end{figure*}

\subsection{Feedback and impact of joints in  rehabilitation exercises}

Since our graph-based approach respects the non-linear structure of the skeleton data, we can investigate the natural topological structure of the body. 
Besides, spatial information is extracted via attention-guided graph convolution.
The body's joint roles can be quantified in order to evaluate rehabilitation exercises.

Our ConGRU output does not provide any structural information. Using the adjacency matrix and element-wise multiplication, we inject the graph structure.
Thus we obtain the self-attention map, $M_1$, which shows the attention weights for each body joint with its neighbors in each row. 
The joint role, $\chi^t$ is computed by the column-wise summation over $M_1$.

\begin{figure*}[ht]
\centering
\includegraphics[width=12cm,height=7cm]{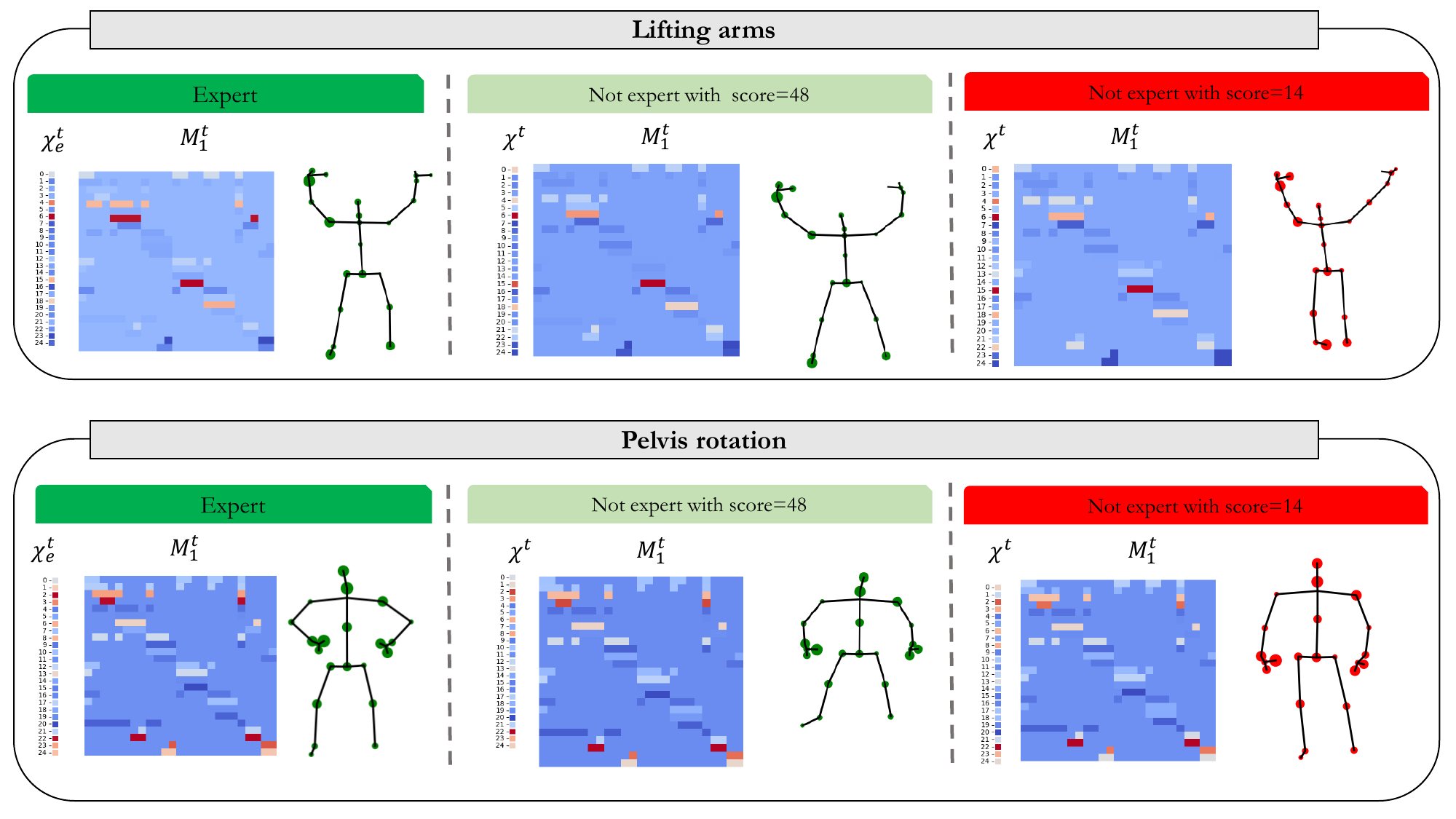}
\caption{ Feedback visualization for different user profiles: expert, not an expert with a good score, and not an expert with a low score.  $\chi^t$ represents the joint role vector and $M_1^t$ the self-attention map (hot colors represent high values).
Colored circles on the skeleton bodies allow the visualization of the attention maps and the role of body joints for different exercises. The larger circle represents the higher role of that joint.}
\label{fig:ExpertNoexpertJointsVisu}
\end{figure*}

Different self-attention maps which emphasize the function of the body's joints are shown in Figure ~\ref{fig:ExpertNoexpertJointsVisu}. The higher emphasis on these joints is evident from the higher attention value taken from the  $\chi^t$.
Even though some joints may receive equal values, they can influence both high and low trial ratings.
However, some joints play a bigger role in determining low ratings. The patient needs to concentrate on joints like these.

Computing joint's role for not expert users, as illustrated in Figure ~\ref{fig:ExpertNoexpertJointsVisu}, shows that they differ from the expert's pattern when the patient receives a low assessment score ($<20$). In this visualization, we display both lifting arms (Ex1), where users are mainly moving their arms, and Pelvis rotation (Ex4), where users are making subtle rotations with the most stress on the spinal column.

In Figure ~\ref{fig:ExercisesJointRoleVisu}, we display the impact of various joints on specific KIMORE dataset actions. More specifically, using the attention values given by $\chi$, we visualize in this figure how the role of joints varies according to different rehabilitation exercises. We can notice that in the first exercise (lifting arm), joints that play an important role are: the thumb, elbow, wrist, and spine (hot colors). The same is observed in exercise (pelvis rotation), major contributing joints are the wrist and spine.

\section{Conclusion}\label{sec5}

The paper introduces a proposed attention-based D-STGCNT model for evaluating physical rehab exercises. The model takes 3D skeleton movement data in graph form as input and outputs a score evaluating the quality of the executed exercise. An extended architecture of the popular STGCN was proposed with transformers. Our extended STGCN architecture employs first, dense connections to learn complex features and patterns. Furthermore, ConvGRU layers utilize a self-attention mechanism on the adjacency matrix of body joints, acknowledging the fact that each body joint holds a different level of significance in exercise evaluation. Analysis of attention values enables to determine the key body joints that greatly affect the final score, thereby giving users insight to enhance their performance in future attempts. Additionally, by employing transformers our model overcomes LSTMs limitations by efficiently processing sequential data with variable-length inputs. This is important in 3D skeleton exercise assessment, as the number of joints and frames in a given action can vary. 
The proposed model outperforms quantitatively state-of-the-art results on both KIMORE and UI-PRMD datasets. Qualitative illustrations and a feedback regarding the importance of joints are given and commented. In future works, we intend to investigate continuous assessment and develop a visual feedback through a graphical interface.

\bibliographystyle{cas-model2-names}
\bibliography{refs}

\end{document}